\documentclass[12pt]{article}
\usepackage{amsmath}
\usepackage{graphicx}
\usepackage{bm}
\usepackage{fancyhdr}
\usepackage{amssymb}
\usepackage[all]{xy}
\renewcommand{\theequation}{\arabic{section}.\arabic{equation}}

\begin{document}



\def\a{\alpha}
\def\b{\beta}
\def\d{\delta}
\def\e{\epsilon}
\def\g{\gamma}
\def\h{\mathfrak{h}}
\def\k{\kappa}
\def\l{\lambda}
\def\o{\omega}
\def\p{\wp}
\def\r{\rho}
\def\t{\tau}
\def\s{\sigma}
\def\z{\zeta}
\def\x{\xi}
\def\V={{{\bf\rm{V}}}}
 \def\A{{\cal{A}}}
 \def\B{{\cal{B}}}
 \def\C{{\cal{C}}}
 \def\D{{\cal{D}}}
\def\K{{\cal{K}}}
\def\O{\Omega}
\def\R{\bar{R}}
\def\T{{\cal{T}}}
\def\L{\Lambda}
\def\f{E_{\tau,\eta}(sl_2)}
\def\E{E_{\tau,\eta}(sl_n)}
\def\Zb{\mathbb{Z}}
\def\Cb{\mathbb{C}}

\def\R{\overline{R}}

\def\beq{\begin{equation}}
\def\eeq{\end{equation}}
\def\bea{\begin{eqnarray}}
\def\eea{\end{eqnarray}}
\def\ba{\begin{array}}
\def\ea{\end{array}}
\def\no{\nonumber}
\def\le{\langle}
\def\re{\rangle}
\def\lt{\left}
\def\rt{\right}

\newtheorem{Theorem}{Theorem}
\newtheorem{Definition}{Definition}
\newtheorem{Proposition}{Proposition}
\newtheorem{Lemma}{Lemma}
\newtheorem{Corollary}{Corollary}
\newcommand{\proof}[1]{{\bf Proof. }
        #1\begin{flushright}$\Box$\end{flushright}}

\baselineskip=20pt

\newfont{\elevenmib}{cmmib10 scaled\magstep1}
\newcommand{\preprint}{
   \begin{flushleft}
   \end{flushleft}\vspace{-1.3cm}
   \begin{flushright}\normalsize
   \end{flushright}}
\newcommand{\Title}[1]{{\baselineskip=26pt
   \begin{center} \Large \bf #1 \\ \ \\ \end{center}}}
\newcommand{\Author}{\begin{center}
   \large \bf
Kun Hao${}^{a}$,~Junpeng Cao${}^{b,c}$,~Guang-Liang Li${}^{d}$,~Wen-Li Yang${}^{a,e}\footnote{Corresponding author:
wlyang@nwu.edu.cn}$,
 ~ Kangjie Shi${}^a$ and~Yupeng Wang${}^{b,c}\footnote{Corresponding author: yupeng@iphy.ac.cn}$
 \end{center}}
\newcommand{\Address}{\begin{center}

     ${}^a$Institute of Modern Physics, Northwest University,
     Xian 710069, China\\
     ${}^b$Beijing National Laboratory for Condensed Matter
           Physics, Institute of Physics, Chinese Academy of Sciences, Beijing
           100190, China\\
     ${}^c$Collaborative Innovation Center of Quantum Matter, Beijing,
     China\\
     ${}^d$Department of Applied Physics, Xian Jiaotong University, Xian 710049, China\\
     ${}^e$Beijing Center for Mathematics and Information Interdisciplinary Sciences, Beijing, 100048,  China

   \end{center}}

\preprint \thispagestyle{empty}
\bigskip\bigskip\bigskip

\Title{Exact solution of an
$su(n)$ spin torus} \Author

\Address \vspace{1cm}

\begin{abstract}
The trigonometric $su(n)$ spin chain  with anti-periodic boundary
condition ($su(n)$  spin torus) is demonstrated to be Yang-Baxter
integrable. Based on some intrinsic properties of the $R$-matrix,
certain operator product identities of the transfer matrix are
derived. These identities and the asymptotic behavior of the
transfer matrix together allow us to obtain the exact eigenvalues in
terms of an inhomogeneous $T-Q$  relation via the off-diagonal Bethe
Ansatz.

\vspace{1truecm} \noindent {\it PACS:} 75.10.Pq, 03.65.Vf, 71.10.Pm


\end{abstract}

\newpage



\section{Introduction}
\label{intro} \setcounter{equation}{0}

Study on quantum integrable models \cite{bax1,Kor93} has played an essential role in many areas of physics, such as  condensed matter physics,
quantum field theory, the AdS/CFT \cite{Mal98, Bei12} correspondence in string theory, nuclear physics, atomic and molecular physics and ultracold atoms.
Strikingly, the algebraic Bethe Ansatz (BA) method \cite{Tak79} to solve quantum integrable models with obvious reference states has inspired and led to remarkable developments in different branches of mathematical physics in the past decades.
While for quantum integrable models without $U(1)$-symmetry, obvious reference states are usually absent,
making the conventional Bethe Ansatz methods \cite{bax1,Tak79,Bet31,Skl78,Alc87,Skl88} almost inapplicable.
Recently, a new approach, i.e., the off-diagonal Bethe Ansatz (ODBA)\cite{Cao1} (for comprehensive introduction we refer the reader to \cite{Wan15}) was proposed to obtain exact solutions of generic integrable models either with or without $U(1)$ symmetry.
Several long-standing models were then solved \cite{Cao1,Cao13NB875,Cao14NB886,Cao13NB877,Cao14JHEP143,Li14,Zha13,Hao14} via this method. It should be remarked that some other interesting methods such as
the q-Onsager algebra method \cite{Bas1,Bas2,Bas3}, the modified algebraic Bethe  ansatz method \cite{Bel13,Bel15,Bel15-1,Ava15} and the Sklyanin's separation of variables (SoV) method \cite{Skl92} were also applied to some integrable models  related to the $su(2)$ algebra \cite{Fra08,Nic12,Fad14,Kit14}.

Quantum spin models provide a typical setting of quantum fluctuations leading to various exotic spin liquid states \cite{fazekas74,misguich03}.
The Bethe Ansatz solution \cite{Cao1} of the spin-$1/2$ chain with anti-periodic boundary conditions ($su(2)$ topological spin torus or quantum M\"obius stripe) is a well-known example to reveal topological nature of elementary excitations in such kind of systems. An interesting issue is to study the high-rank systems with topological boundaries.
We note that the models of $su(n)$ quantum spin systems are far from merely theoretical
exercises: It could be realized either in cold-atom systems in
optical lattices \cite{honerkamp04,hofstetter05,buchler05},
in quantum dot arrays \cite{onufriev}, or in spin systems with orbital degrees of freedom \cite{mila2003}. The aim of the present work is to study the integrability  and exact spectrum of the trigonometric $su(n)$ chain with anti-periodic boundary condition with the nested ODBA \cite{Cao14JHEP143}.

The paper is organized as follows. Section 2 serves as an introduction of our notations and some basic ingredients. The commuting transfer matrix associated with the
$su(n)$ spin torus is  constructed to show the integrability of the model.
In section 3, taking the $su(3)$ spin torus as a concrete example, we derive some operator identities based on intrinsic
properties of the $R$-matrix, which allow us to give the eigenvalues of the transfer matrix
in terms of a nested inhomogeneous $T-Q$ relation. The corresponding Bethe Ansatz equations (BAEs) are also given. The generalization
to $su(n)$ case is given in section 4. We summarize our results and give some discussions in Section 5.
The generic integrable twisted boundary conditions are shown  in Appendix A. Some details about the $su(4)$ case, which could be crucial to understand the procedure for $n\geq4$, are given in Appendix B.

\section{$su(n)$ spin torus }
\setcounter{equation}{0}

Let ${\rm\bf V}$ be an $n$-dimensional linear space with an
orthonormal basis $\{|i\rangle|i=1,\cdots,n\}$, we introduce the
Hamiltonian \bea H=\sum_{j=1}^Nh_{j\,j+1},\label{Ham} \eea where $N$
is the number of sites and $h_{j\,j+1}$ is the local Hamiltonian
given by \bea h_{j\,j+1}=\frac{\partial}{\partial
u}\lt.\lt\{P_{j\,j+1}\,R_{j\,j+1}(u)\rt\}\rt|_{u=0}. \eea Here $P$
is the permutation operator on the tensor space ${\rm\bf V}\otimes
{\rm\bf V}$;  the $R$-matrix $R(u)\in {\rm End}({\rm\bf V}\otimes
{\rm\bf V})$ is the trigonometric $R$-matrix associated with the
quantum group \cite{Cha94} $U_q(\widehat{su(n)})$, which was  given in
\cite{Che80,Chu80,Sch81,Bab81,Perk81} and further studied in
\cite{Perk83,Schu83,Perk06,Baz85,Jim86}\footnote{The $R$-matrix
given by (\ref{R-matrix-1}) corresponds to  the so-called principal
gradation, which is related to the $R$-matrix in homogeneous
gradation by some gauge transformation \cite{Nep02}.}
 \bea R(u) &=&
\sinh({u} + \eta) \sum_{k=1}^{n} E^{k\,, k}\otimes E^{k\,, k}+
\sinh{u}
\sum_{k \ne l}^n E^{k\,, k}\otimes E^{l\,, l} \no \\
&&+ \sinh\eta \left(\sum_{k<l}^ne^{\frac{n-2(l-k)}{n}u} +
\sum_{k>l}^n e^{-\frac{n-2(k-l)}{n}u} \right) E^{k\,, l}\otimes
E^{l\,, k} ,\label{R-matrix-1} \eea where the $n^2$ fundamental
matrices $\{E^{k,l}|k,l=1,\cdots,n\}$ are all $n\times n$ matrices
with matrix entries
$(E^{k,l})^{\a}_{\b}=\delta^k_{\a}\,\delta^l_{\b}$ and $\eta$ is the
crossing parameter. The $R$-matrix satisfies the quantum Yang-Baxter
equation (QYBE)
\begin{eqnarray}
 R_{12}(u_1-u_2)R_{13}(u_1-u_3)R_{23}(u_2-u_3)=
 R_{23}(u_2-u_3)R_{13}(u_1-u_3)R_{12}(u_1-u_2), \label{QYB}
\end{eqnarray}
and possesses the properties:
\begin{eqnarray}
 &&\hspace{-1.45cm}\mbox{
 Initial condition}:\hspace{42.5mm}R_{12}(0)= \sinh\eta\, P_{1\,2},\label{Initial}\\[4pt]
 &&\hspace{-1.5cm}\mbox{
 Unitarity}:\hspace{11.5mm}R_{12}(u)R_{21}(-u)= \rho_1(u)\times {\rm id},\quad \rho_1(u)=-\sinh(u+\eta)\sinh(u-\eta),\label{Unitarity}\\[4pt]
 &&\hspace{-1.5cm}\mbox{
 Crossing-unitarity}:\,
 R^{t_1}_{12}(u)R_{21}^{t_1}(-u-n\eta)
 =\rho_2(u)\times \mbox{id},\; \rho_2(u)=-\sinh u\sinh(u+n\eta),
 \label{crosing-unitarity}\\[4pt]
 &&\hspace{-1.4cm}\mbox{Fusion conditions}:\hspace{22.5mm}\, R_{12}(-\eta)\propto P^{(-)}_{1\,2},\label{Fusion}\\[4pt]
 &&\hspace{-1.4cm}\mbox{Periodicity}:\hspace{22.5mm}\, R_{12}(u+i\pi)=-h_1\,R_{12}(u)\,h_1^{-1}=-h_2^{-1}\,R_{12}(u)\,h_2
 .\label{Periodicity-R}
\end{eqnarray}
Here $R_{21}(u)=P_{1\,2}R_{12}(u)P_{1\,2}$;
$P^{(-)}_{1\,2}$  is the q-deformed anti-symmetric
 project operator in the tensor product space  ${\rm\bf
V} \otimes {\rm\bf V} $ (such as below (\ref{P-3}) and (\ref{Fusion-3}));  $t_i$ denotes the transposition in the
$i$-th space; $h$ is an $n\times n$ diagonal matrix given by
\bea
h=\lt(\begin{array}{cccc}1&&&\\
&\omega_n&&\\
&&\ddots&\\
&&&\omega_n^{n-1}\\
\end{array}\right),\quad \omega_n=e^{\frac{2i\pi}{n}},\quad {\rm and}\quad  h^n=1. \label{h-matrix}
\eea
Here and below we adopt the standard notation: for any
matrix $A\in {\rm End}({\rm\bf V})$, $A_j$ is an embedding operator
in the tensor space ${\rm\bf V}\otimes {\rm\bf V}\otimes\cdots$,
which acts as $A$ on the $j$-th space and as an identity on the
other factor spaces; $R_{ij}(u)$ is an embedding operator of
$R$-matrix in the tensor space, which acts as an identity on the
factor spaces except for the $i$-th and $j$-th ones.

In order to construct the quantum spin chain with integrable twisted boundary condition \cite{deV84}, let us introduce an $n\times n$ twist matrix $g$
\bea
g=\lt(\begin{array}{cccc}&&&1\\
1&&&\\
&\ddots&&\\
&&1&\\
\end{array}\right),\quad g^n=1.\label{g-matrix}
\eea
It can be easily checked that the $R$-matrix (\ref{R-matrix-1}) is invariant with $g$, namely,
\bea
&&R_{12}(u)\,g_1\,g_2=g_1\,g_2\,R_{12}(u),\label{Invariant-R}\\
&&h\,g=\omega_n\,g\,h,\label{hg-relation}
\eea
where $h$ is given by (\ref{h-matrix}). (Generic twist matrix satisfing the above equation
is given in Appendix A.) This property enables us to construct the integrable $su(n)$ spin torus model.

Similar to the $su(2)$ spin torus (or the XXZ spin chain with anti-periodic boundary condition) \cite{Bat95},
the $su(n)$ spin torus is described by the Hamiltonian $H$ given by (\ref{Ham}) with  anti-periodic
boundary conditions
\bea
E^{k,l}_{N+1}=g_1\,E^{k,l}_1\,g_1^{-1},\quad k,l=1,\cdots,n.\label{anti-boundary}
\eea
Let us introduce the ``row-to-row" monodromy matrix
$T(u)$, an $n\times n$ matrix with operator-valued elements
acting on ${\rm\bf V}^{\otimes N}$, \bea T_0(u)
=R_{0N}(u-\theta_N)R_{0\,N-1}(u-\theta_{N-1})\cdots
R_{01}(u-\theta_1).\label{Monodromy-1} \eea Here
$\{\theta_j|j=1,\cdots,N\}$ are generic free complex parameters
which are usually called as inhomogeneity parameters.
The transfer matrix $t(u)$ of the associated spin chain describing the Hamiltonian (\ref{Ham}) with the antiperiodic
boundary condition (\ref{anti-boundary}) can be constructed similarly as  \cite{Skl88,deV84,Bat95}
\begin{eqnarray}
t(u)&=&tr_0\lt\{g_0\,T_0(u)\rt\}.\label{transfer}
\end{eqnarray}
The QYBE (\ref{QYB}) and the definition (\ref{Monodromy-1}) of the monodromy matrix $T(u)$ imply that the matrix elements of $T(u)$
satisfy the Yang-Baxter algebra:
\bea
R_{12}(u-v)\,T_1(u)\,T_2(v)=T_2(v)\,T_1(u)\,R_{12}(u-v).\label{Yang-Baxter-algebra}
\eea
The above relation and  the invariant relation (\ref{Invariant-R})
lead to the fact that the transfer matrices $t(u)$ given by (\ref{transfer}) with
different spectral parameters are mutually commuting:
$[t(u),t(v)]=0$. The  Hamiltonian (\ref{Ham}) with the
anti-periodic boundary condition (\ref{anti-boundary})
can be obtained from the transfer matrix as
\begin{eqnarray}
&&H=\sinh\eta\, \frac{\partial \ln t(u)}{\partial u}|_{u=0,\{\theta_j\}=0}.
\end{eqnarray}
This ensures the integrability of the $su(n)$ spin torus.
The aim of this paper is to obtain  eigenvalues of the transfer matrix $t(u)$ specified by the twist matrix (\ref{g-matrix}) via ODBA.

\section{ODBA solution of the $su(3)$ spin torus}
\setcounter{equation}{0}

Following the method developed in \cite{Cao14JHEP143}, we apply the fusion techniques \cite{Kar79,Kulish81,Kulish82,Kirillov86} to study the $su(n)$ spin torus. For this purpose, besides the fundamental transfer
matrix $t(u)$ some other fused transfer matrices $\{t_j(u)|j=1,\cdots,n\}$, which commute with each other and include
the original one as  $t_1(u)=t(u)$, should be constructed through an anti-symmetric fusion procedure.
In this section, we present the results for the $su(3)$ spin torus.

\subsection{Operator identities of the transfer matrices}
For the $su(3)$ case, the $R$-matrix $R(u)$ given by (\ref{R-matrix-1}) reads
\bea\label{R-matrix-su3q}
R(u)=
\left(
  \begin{array}{ccc|ccc|ccc}
 \bar a(u) &    &    &    &    &    &    &    &   \\
      &\bar b(u)&    &\bar c(u)&    &    &    &    &   \\
      &    &\bar b(u)&    &    &    &\bar d(u)&    &   \\
\hline
      &\bar d(u)&    &\bar b(u)&    &    &    &    &   \\
      &    &    &    &\bar a(u)&    &    &    &   \\
      &    &    &    &    &\bar b(u)&    &\bar c(u)&   \\
\hline
      &    &\bar c(u)&    &    &    &\bar b(u)&    &    \\
      &    &    &    &    &\bar d(u)&    &\bar b(u)&    \\
      &    &    &    &    &    &    &    &\bar a(u)\\
\end{array}
\right)
\eea
with
\bea
&\displaystyle \bar a(u)=\sinh(u+\eta),\quad &\bar b(u)=\sinh(u),\no\\[4pt]
&\displaystyle \bar c(u)=e^{{u\over3}}\sinh(\eta),\quad &\bar d(u)=e^{-{u\over3}}\sinh(\eta),\no
\eea
and the twist matrix  $g$ given by (\ref{g-matrix}) becomes
\bea
g=\lt(\begin{array}{ccc}0&0&1\\
1&0&0\\
0&1&0
\end{array}\right),\quad {\rm and}\,\,g^3=1.\label{g-matrix-3}
\eea
Following \cite{Hao14}, let us introduce the following vectors in the tensor space ${\rm\bf V}\otimes{\rm\bf V}$ associated
with the $R$-matrix (\ref{R-matrix-su3q})
\bea
|\Phi^{(1)}_{12}\rangle&=&\frac{1}{\sqrt{2e^{-\eta\over3}\cosh({\eta\over3})}}(|1,2\rangle-e^{-{\eta\over3}}|2,1\rangle),\no\\[6pt]
|\Phi^{(2)}_{12}\rangle&=&\frac{1}{\sqrt{2e^{\eta\over3}\cosh({\eta\over3})}}(|1,3\rangle-e^{{\eta\over3}}|3,1\rangle),\no\\[6pt]
|\Phi^{(3)}_{12}\rangle&=&\frac{1}{\sqrt{2e^{-\eta\over3}\cosh({\eta\over3})}}(|2,3\rangle-e^{-{\eta\over3}}|3,2\rangle),
\eea
and a vector in the tensor space ${\rm\bf V}\otimes{\rm\bf V}\otimes{\rm\bf V}$,
\bea
|\Phi_{123}\rangle&=&\frac{1}{\sqrt{6e^{-{\eta\over3}}\cosh({\eta\over3})}} (|1,2,3\rangle-e^{-{\eta\over3}}|1,3,2\rangle-e^{-{\eta\over3}}|2,1,3\rangle\no\\[6pt]
&&+|2,3,1\rangle+|3,1,2\rangle-e^{-{\eta\over3}}|3,2,1\rangle).
\eea
From these vectors we can construct the associated projectors\footnote{These operators are $q$-deformed anti-symmetric projectors and in contrast to the rational ones,  $P^{(-)}_{21}=P_{12}P^{(-)}_{12}P_{12}\neq P^{(-)}_{12}$.}
\bea
P^{(-)}_{12}&=&|\Phi^{(1)}_{12}\rangle\langle\Phi^{(1)}_{12}|
+|\Phi^{(2)}_{12}\rangle\langle\Phi^{(2)}_{12}|+|\Phi^{(3)}_{12}\rangle\langle\Phi^{(3)}_{12}|,\label{P-3}\\[6pt]
P^{(-)}_{123}&=&|\Phi_{123}\rangle\langle\Phi_{123}|.
\eea
Direct calculation shows that the $R$-matrix given by (\ref{R-matrix-su3q}) at some degenerate points is proportional to the projectors,
\bea
R_{12}(-\eta)=P^{(-)}_{12}\times S^{(-)}_{12},\qquad
R_{12}(-\eta)R_{13}(-2\eta)R_{23}(-\eta)=P^{(-)}_{123}\times S^{(-)}_{123},\label{Fusion-3}
\eea
where $S^{(-)}_{12}\in {\rm End}({\rm\bf V}\otimes{\rm\bf V})$ and $S^{(-)}_{123}\in {\rm End}({\rm\bf V}\otimes{\rm\bf V}\otimes{\rm\bf V})$ are some non-degenerate diagonal matrices.

Now we are in position to derive some operator product identities of the transfer matrices which are  crucial to obtain eigenvalues of the transfer matrix. Let us evaluate  the products of the monodromy matrices at some special points, which lead to the  useful relation (for details we refer the readers to the book \cite{Wan15}, chapter 7),
\bea
T_1(\theta_j)T_2(\theta_j-\eta)&=&P_{21}^{(-)}T_1(\theta_j)T_2(\theta_j-\eta),\label{Fusion-Products-1}\\[6pt]
T_1(\theta_j)P^{(-)}_{32}T_2(\theta_j-\eta)T_3(\theta_j-2\eta)P^{(-)}_{32}&=&
P^{(-)}_{321}T_1(\theta_j)T_2(\theta_j-\eta)T_3(\theta_j-2\eta)P^{(-)}_{32}.\label{Fusion-Products-2}
\eea
The invariance (\ref{Invariant-R}) of the $R$-matrix $R(u)$ and the relations (\ref{Fusion-3}) imply the relations
\bea
[g_1\,g_2,\,P^{(-)}_{21}]=0=[g_1\,g_2\,g_3,\,P^{(-)}_{321}].\label{comm-g-pm}
\eea
With the help of  the above relations (\ref{Fusion-Products-1})-(\ref{comm-g-pm}), we can calculate the products of the
fundamental transfer matrices at some special points
\bea
t(\theta_j)t(\theta_j-\eta)&=&tr_{12}\{g_1T_1(\theta_j)g_2T_2(\theta_j-\eta)\}\no\\[4pt]
&=&tr_{12}\{g_1g_2T_1(\theta_j)T_2(\theta_j-\eta)\}\no\\[4pt]
&=&tr_{12}\{g_1g_2P_{21}^{(-)}T_1(\theta_j)T_2(\theta_j-\eta)\}\no\\[4pt]
&\overset{(\ref{comm-g-pm})}{=}&tr_{12}\{P_{21}^{(-)}g_1g_2P_{21}^{(-)}P_{21}^{(-)}T_1(\theta_j)T_2(\theta_j-\eta)P_{21}^{(-)}\}\no\\[4pt]
&=&tr_{12}\{g_{<12>}T_{<12>}(\theta_j)\}\no\\[4pt]
&=&t_2(\theta_j),\label{trans-trans}
\eea
where
\bea
g_{<12>}\equiv P_{21}^{(-)}g_1g_2P_{21}^{(-)},\quad\quad T_{<12>}(u)\equiv P_{21}^{(-)}T_1(u)T_2(u-\eta)P_{21}^{(-)},\no\eea
and the fused transfer matrix $t_2(u)$ is given by
\bea
t_2(u)=tr_{12}\{g_{<12>}T_{<12>}(u)\}.\label{transfer-2}
\eea
Moreover, we can derive
\bea
t(\theta_j)t_2(\theta_j-\eta)&=&tr_{123}\{g_1T_1(\theta_j)P^{(-)}_{32}g_2g_3P^{(-)}_{32}P^{(-)}_{32}
T_2(\theta_j-\eta)T_3(\theta_j-2\eta)\}\no\\[4pt]
&=&tr_{123}\{g_1g_{<23>}T_1(\theta_j)P^{(-)}_{32}T_2(\theta_j-\eta)T_3(\theta_j-2\eta)P^{(-)}_{32}\}\no\\[4pt]
&=&tr_{123}\{g_1g_2g_3T_1(\theta_j)T_{<23>}(\theta_j-\eta)\}\no\\
&\overset{(\ref{Fusion-Products-2})}{=}&tr_{123}\{g_1g_2g_3 P^{(-)}_{321}T_1(\theta_j)T_2(\theta_j-\eta)
T_3(\theta_j-2\eta)P^{(-)}_{32}\}\no\\[4pt]
&=&tr_{123}\{g_{<123>}T_{<123>}(\theta_j)\}\no\\[4pt]
&=&t_3(\theta_j),\label{trans-trans2}
\eea
where
\bea
g_{<123>}&=&P^{(-)}_{321}g_1g_2g_3P^{(-)}_{321},\no\\[6pt]
T_{<123>}(u)&=&P^{(-)}_{321}T_1(u)T_2(u-\eta)T_3(u-2\eta)P^{(-)}_{321}\no\\[6pt]
&=&\prod^N_{l=1}\sinh(u-\theta_l+\eta)\sinh(u-\theta_l-\eta)
\sinh(u-\theta_l-2\eta)P^{(-)}_{321}.\no
\eea
Direct calculation shows that
\bea
t_3(u)=tr_{123}\{g_{<123>}T_{<123>}(u)\}=(-1)^{(3-1)}{\rm Det}_{q}T(u)\times{\rm id},
\eea
where the quantum determinant function ${\rm Det}_{q}T(u)$ reads
\bea
{\rm Det}_{q}T(u)=\prod^N_{l=1}\sinh(u-\theta_l+\eta)\sinh(u-\theta_l-\eta)
\sinh(u-\theta_l-2\eta).\no
\eea
Then the relation (\ref{trans-trans2}) becomes
\bea
t(\theta_j)t_2(\theta_j-\eta)={\rm Det}_{q}T(\theta_j)\times{\rm id}.\label{trans-trans-1}
\eea
Using the initial condition (\ref{Initial}) and the unitarity relation (\ref{Unitarity}),
we can derive that the fused transfer matrix $t_2(u)$ vanishes at the points: $\theta_j+\eta$, i.e.,
\bea
t_2(\theta_j+\eta)=0,\quad j=1,\cdots,N.
\label{fusion-orth}\eea
Moreover, it follows from the fusion procedure and the QYBE (\ref{QYB}) that the fused transfer matrices constitute commutative families, namely,
\bea
[t_i(u),t_j(v)]=0,\qquad i,j=1,2,3.\label{commu-trans-su3}
\eea

Now let us consider the periodicities and the asymptotic behaviors of the transfer matrices $t(u)$ and $t_2(u)$. The periodicity (\ref{Periodicity-R}) of the
the $R$-matrix $R(u)$ and the definition (\ref{Monodromy-1}) of the monodromy matrix give rise to the relation
\bea
T(u+i\pi)=(-1)^N\,h\,T(u)\,h^{-1},\quad h=\left(
   \begin{array}{ccc}
    1 &  &  \\
      & \omega_3 &  \\
      &  & \omega_3^2 \\
   \end{array}
 \right), \label{Periodicity-T}
\eea
with $\omega_3=e^{\frac{2i\pi}{3}}$. Keeping the fact that $gh=\omega_3 gh$ and using the relation (\ref{Periodicity-T}), we can derive that
the transfer matrices $t(u)$ and $t_2(u)$ satisfy the periodicities:
\bea
t(u+i\pi)=(-1)^N\,e^{-\frac{2i\pi}{3}}t(u), \quad t_2(u+i\pi)=e^{-\frac{4i\pi}{3}}t_2(u).\label{Periodicity-T-1}
\eea
The explicit expression (\ref{R-matrix-su3q}) of the $R$-matrix and the definitions (\ref{Monodromy-1}), (\ref{transfer}) and (\ref{transfer-2}) allow us to derive that $e^{-\frac{u}{3}}t(u)$ and $e^{\frac{u}{3}}t_2(u)$, as functions of $u$, are polynomials of $e^{\pm u}$ with the asymptotic behaviors:
\bea
e^{-\frac{u}{3}}t(u)&\propto& e^{\pm(N-1)u}+\cdots,\quad u\rightarrow\pm\infty,\label{transfer-asym}\\[6pt]
e^{\frac{u}{3}}t_2(u)&\propto& e^{\pm(2N-1)u}+\cdots,\quad u\rightarrow\pm\infty.\label{transfer-asym-1}
\eea

\subsection{Inhomogeneous $T-Q$ relation and the associated BAEs}

The commutativity (\ref{commu-trans-su3}) of the transfer matrices $t(u)$ and $t_2(u)$ with different spectral parameters implies that they have common eigenstates.
Let $|\Psi\rangle$ be a common eigenstate of $\{t_m(u)\}$, which dose not depend upon $u$, with the eigenvalues $\Lambda_m(u)$,
\bea
t_m(u)|\Psi\rangle=\Lambda_m(u)|\Psi\rangle,\qquad m=1,2, 3.\no
\eea
The fusion relations (\ref{trans-trans}), (\ref{trans-trans-1}) and (\ref{fusion-orth}) imply that the eigenvalues $\L_i(u)$ satisfy the relations
\bea
&&\Lambda(\theta_j)\Lambda_m(\theta_j-{\eta})=\Lambda_{m+1}(\theta_j),~~~~~ m=1,2,\quad j=1,\cdots,N,\label{Eigenvalue-function-1}\\[4pt]
&&\Lambda_3(u)=\prod^N_{l=1}\sinh(u-\theta_l+{\eta})\prod^{2}_{k=1}\sinh(u-\theta_l-k{\eta}),
\label{Eigenvalue-function-2}\\[4pt]
&&\Lambda_2(\theta_j+{\eta})=0,~~~~~~j=1,\cdots,N.\label{Eigenvalue-function-3}
\eea
The periodicity properties (\ref{Periodicity-T-1}) of the transfer matrices enable us to derive that the eigenvalues $\L_i(u)$ satisfy the associated
periodicity relations
\bea
\Lambda(u+i\pi)=e^{-\frac{2i\pi}{3}}(-1)^N\Lambda(u),\quad \Lambda_2(u+i\pi)=e^{-\frac{4i\pi}{3}}\Lambda_2(u).\label{Eigenvalue-function-4}
\eea
In addition, the asymptotic behaviors (\ref{transfer-asym})-(\ref{transfer-asym-1}) of the transfer matrices and their definitions (\ref{transfer})
and (\ref{transfer-2}) lead to the fact that the eigenvalues $\L_i(u)$, as a function of $u$, can be expressed as
\bea
\Lambda(u)&=&e^{\frac{u}{3}}\lt\{I^{(1)}_1e^{(N-1)u}+I^{(1)}_2e^{(N-3)u}+\cdots+I^{(1)}_Ne^{-(N-1)u}\rt\},\label{asym-su3-1}\\[6pt]
\Lambda_2(u)&=&e^{-\frac{u}{3}}\lt\{I^{(2)}_1e^{(2N-1)u}+I^{(2)}_2e^{(2N-3)u}+\cdots+I^{(2)}_{2N}e^{-(2N-1)u}\rt\},\label{asym-su3-2}
\eea
where $\{I^{(1)}_j|j=1,\cdots,N\}$ and $\{I^{(2)}_j|j=1,\cdots,2N\}$ are $3N$ constants which are eigenstate dependent. Then these constants can be
completely determined by the $3N$ equations (\ref{Eigenvalue-function-1})-(\ref{Eigenvalue-function-3}). The above relations
(\ref{Eigenvalue-function-1})-(\ref{asym-su3-2}) allow us to express the eigenvalues
$\L_i(u)$ in terms of some inhomogeneous $T-Q$ relations \cite{Wan15}.

For this purpose, let us introduce some functions:
\bea
&&a(u)=\prod_{l=1}^N\sinh(u-\theta_l+\eta),\quad d(u)= \prod_{l=1}^N\sinh(u-\theta_l)=a(u-\eta),\label{ad-functions}\\[4pt]
&&Q^{(i)}(u)=\prod_{l=1}^{N_i}\sinh(u-\l^{(i)}_l),\quad i=1,2,3,4,\no\\[4pt]
&&f_1(u)=f_1^{(+)}e^u+f_1^{(-)}e^{-u},\quad f_2(u)=f_2^{(-)}e^{-u},\no
\eea
where the $(N_1+N_2+N_3+N_4)$ parameters $\{\l^{(i)}_l|l=1,\cdots, N_i;\,i=1,2,3,4\}$ and the parameters $f_1^{(\pm)}$ and $f_2^{(-)}$ will be specified later by the associated
BAEs (\ref{BAE-1})-(\ref{BAE-8}). For convenience, we introduce further the following notations:
\bea
&&Z_1(u)=e^{\phi_1}e^{\frac{4u}{3}}a(u)\frac{Q^{(1)}(u-\eta)}{Q^{(2)}(u)},\no\\[4pt]
&&Z_2(u)=e^{-\phi_1}\omega_3 e^{-{2(u+\eta)\over3}} d(u)\frac{Q^{(2)}(u+\eta)Q^{(3)}(u-\eta)}{Q^{(1)}(u)Q^{(4)}(u)},\no\\[4pt]
&&Z_3(u)=\omega_3^2 e^{-{2(u+2\eta)\over3}} d(u)\frac{Q^{(4)}(u+\eta)}{Q^{(3)}(u)},\no\\[4pt]
&&X_1(u)=e^{\frac{u}{3}}a(u)d(u)\frac{Q^{(3)}(u-\eta)f_1(u)}{Q^{(1)}(u)Q^{(2)}(u)},\no\\[4pt]
&&X_2(u)=e^{\frac{u}{3}}a(u)d(u)\frac{Q^{(2)}(u+\eta)f_2(u)}{Q^{(3)}(u)Q^{(4)}(u)},
\eea
where $\phi_1$ is a parameter to be determined later.  The eigenvalues $\{\Lambda_i(u)\}$ satisfing
(\ref{Eigenvalue-function-1})-(\ref{Eigenvalue-function-3}) and (\ref{asym-su3-1})-(\ref{asym-su3-2})
can be expressed in terms of  the inhomogeneous $T-Q$ relations as follows
\footnote{It is well-known that for the $su(n)$ spin chain  with the periodic boundary condition the eigenvalues are given by
the  usual homogeneous $T-Q$ relations which  only have $n-1$ types of $Q$-functions. However, it seems
that for the anti-periodic boundary condition case the associated inhomogeneous $T-Q$ relations (\ref{T-Q-1})-(\ref{T-Q-2})
(or (\ref{T-Q-n-1}) for the generic $n$) have to involve more types of $Q$-functions (c.f. $2(n-1)$ types of $Q$-functions see (\ref{T-Q-n-1}) below)
than that of the periodic case, except the $n=2$ case \cite{Cao1}.
}
\bea
&&\hspace{-1.2truecm}\L(u)=Z_1(u)+Z_2(u)+Z_3(u)+X_1(u)+X_2(u)\no\\[6pt]
&&\hspace{-1.2truecm}~~~~~~=e^{\frac{u}{3}}\lt\{e^{\phi_1}e^u a(u)\frac{Q^{(1)}(u-\eta)}{Q^{(2)}(u)}
       +e^{-\phi_1}\omega_3 e^{-u-{2\eta\over3}} d(u)\frac{Q^{(2)}(u+\eta)Q^{(3)}(u-\eta)}{Q^{(1)}(u)Q^{(4)}(u)}\rt.\no\\[4pt]
&&\hspace{-1.2truecm}\quad\quad\quad+\omega_3^2 e^{-u-{4\eta\over3}} d(u)\frac{Q^{(4)}(u+\eta)}{Q^{(3)}(u)}
       +a(u)d(u)\frac{Q^{(3)}(u-\eta)f_1(u)}{Q^{(1)}(u)Q^{(2)}(u)}\no\\[4pt]
&&\hspace{-1.2truecm}\quad\quad\quad
       +a(u)d(u)\lt.\frac{Q^{(2)}(u+\eta)f_2(u)}{Q^{(3)}(u)Q^{(4)}(u)} \rt\},\label{T-Q-1}\\[4pt]
&&\hspace{-1.2truecm}\L_2(u)=Z_1(u)Z_2^{(1)}(u)+Z_1(u)Z_3^{(1)}(u)+Z_2(u)Z_3^{(1)}(u)+X_1(u)Z_3^{(1)}(u)+Z_1(u)X_2^{(1)}(u)\no\\[6pt]
&&\hspace{-1.2truecm}~~~~~~~=e^{-\frac{u}{3}}d(u-\eta)\lt\{\omega_3 e^{u} a(u)\frac{Q^{(3)}(u-2\eta)}{Q^{(4)}(u-\eta)}
       +e^{-\phi_1}e^{-4\eta\over3} e^{-u} d(u)\frac{Q^{(2)}(u+\eta)}{Q^{(1)}(u)}\rt.\no\\[4pt]
&&\hspace{-1.2truecm}\quad\quad\quad+\omega_3^2 e^{\phi_1}e^{-2\eta\over3} e^{u} a(u)\frac{Q^{(1)}(u-\eta)Q^{(4)}(u)}{Q^{(2)}(u)Q^{(3)}(u-\eta)}
       +\omega_3^2e^{-2\eta\over3}\hspace{-0.1truecm}a(u)d(u)\frac{Q^{(4)}(u)f_1(u)}{Q^{(1)}(u)Q^{(2)}(u)}\no\\[4pt]
&&\hspace{-1.2truecm}\quad\quad\quad+e^{\phi_1}\hspace{-0.1truecm}e^{2u-{\eta\over3}}a(u)d(u)
       \lt.\frac{Q^{(1)}(u-\eta)f_2(u-\eta)}{Q^{(3)}(u-\eta)Q^{(4)}(u-\eta)} \rt\}\hspace{-0.1truecm},\label{T-Q-2}\\[4pt]
&&\hspace{-1.2truecm}\L_3(u)=Z_1(u)\,Z_2^{(1)}(u)\,Z_3^{(2)}(u)\no\\[6pt]
&&\hspace{-1.2truecm}~~~~~~~~=\omega_3^3 a(u)d(u-\eta)d(u-2\eta)=a(u)d(u-\eta)d(u-2\eta).
\eea
Here and below we adopt the conventions
\bea
Z_i^{(l)}(u)=Z_i(u-l\eta),\quad X_i^{(l)}(u)=X_i(u-l\eta).\label{Notation}
\eea

To make the inhomogeneous $T-Q$ relations (\ref{T-Q-1}) and (\ref{T-Q-2}) to fulfill the asymptotic behavior
(\ref{asym-su3-1})-(\ref{asym-su3-2}),
the non-negative integers $N_i$ should satisfy the relations
\bea
N_1=N_2=N_3=N_4=N,\label{N-su(3)}
\eea
and the $4N+4$ parameters $\{\l^{(i)}_l|l=1,\cdots, N;\,i=1,2,3,4\}$, $f_1^{(\pm)}$, $f_2^{(-)}$ and $e^{\phi_1}$ satisfy the associated
BAEs\footnote{It is still an interesting open problem to investigate the structure of the Bethe roots of the BAEs (\ref{BAE-1})-(\ref{BAE-8}) 
for a large $N$. One promising strategy might be to study the corresponding elliptical model for the large sites with the crossing parameter 
taken some special values (which will become dense in the whole complex plan when $N\rightarrow\infty$) for which the inhomogeneous $T-Q$ relation reduce to the usual one \cite{Cao14NB886}. This allows one to study the pattern of the corresponding Bethe roots for the large $N$ and then taking the trigonometric limit we can obtain the pattern of the Bethe root for the trigonometric models for a large $N$. This strategy has proven to be very successful for the studying the thermodynamics of the spin-$\frac{1}{2}$ open XXZ chain \cite{Li14}.}:
\bea
&&\omega_3 e^{-\phi_1}e^{-\lambda^{(1)}_j-{2\eta\over3}} \frac{Q^{(2)}(\lambda^{(1)}_j+\eta)}{Q^{(4)}(\lambda^{(1)}_j)}
+a(\lambda^{(1)}_j)\frac{f_1(\lambda^{(1)}_j)}{Q^{(2)}(\lambda^{(1)}_j)}=0,\quad j=1,\cdots,N,\label{BAE-1}\\[4pt]
&&e^{\phi_1}e^{\lambda^{(2)}_j}Q^{(1)}(\lambda^{(2)}_j-\eta)
+d(\lambda^{(2)}_j)
\frac{Q^{(3)}(\lambda^{(2)}_j-\eta)f_1(\lambda^{(2)}_j)}{Q^{(1)}(\lambda^{(2)}_j)}=0,
\quad j=1,\cdots,N,\label{BAE-2}\\[4pt]
&&\omega_3^2 e^{-\lambda^{(3)}_j-{4\eta\over3}}Q^{(4)}(\lambda^{(3)}_j+\eta)
+a(\lambda^{(3)}_j)
\frac{Q^{(2)}(\lambda^{(3)}_j+\eta)f_2(\lambda^{(3)}_j)}{Q^{(4)}(\lambda^{(3)}_j)}=0,
\,\, j=1,\cdots,N,\label{BAE-3}\\[4pt]
&&\omega_3
e^{-\phi_1}e^{-\lambda^{(4)}_j-{2\eta\over3}}\frac{Q^{(3)}(\lambda^{(4)}_j-\eta)}{Q^{(1)}(\lambda^{(4)}_j)}
+a(\lambda^{(4)}_j)\frac{f_2(\lambda^{(4)}_j)}{Q^{(3)}(\lambda^{(4)}_j)}=0,
\quad j=1,\cdots,N,\label{BAE-4} \eea \bea
&&\hspace{-1.2truecm}e^{\phi_1}e^{-\Theta-\chi^{(1)}+\chi^{(2)}}+e^{-2\Theta+\chi^{(1)}+\chi^{(2)}-\chi^{(3)}}f^{(+)}_1=0,\label{BAE-5}\\[4pt]
&&\hspace{-1.2truecm}\omega_3 e^{-\phi_1}e^{-{2\eta\over3}+\Theta-\chi^{(1)}+\chi^{(2)}+\chi^{(3)}-\chi^{(4)}}
+\omega_3^2 e^{-{4\eta\over3}+\Theta-\chi^{(3)}+\chi^{(4)}-N\eta}\no\\[4pt]
&&+e^{2\Theta-N\eta}\lt\{e^{-\chi^{(1)}-\chi^{(2)}+\chi^{(3)}+N\eta}f^{(-)}_1
+e^{+\chi^{(2)}-\chi^{(3)}-\chi^{(4)}-N\eta}f^{(-)}_2\rt\}=0,\label{BAE-6}\\[4pt]
&&\hspace{-1.2truecm}\omega_3 e^{-\Theta-\chi^{(3)}+\chi^{(4)}}+\omega_3^2 e^{\phi_1}e^{-{2\eta\over3}
-\Theta-\chi^{(1)}+\chi^{(2)}+\chi^{(3)}-\chi^{(4)}+N\eta}\no\\[4pt]
&&+e^{-2\Theta+N\eta}\lt\{\omega_3^2 e^{-{2\eta\over3}+\chi^{(1)}+\chi^{(2)}-\chi^{(4)}}f^{(+)}_1
+e^{\phi_1}e^{{2\eta\over3}-\chi^{(1)}+\chi^{(3)}+\chi^{(4)}+N\eta}f^{(-)}_2\rt\}=0,\label{BAE-7}\\[4pt]
&&\hspace{-1.2truecm}e^{-\phi_1}e^{-{4\eta\over3}+\Theta-\chi^{(1)}+\chi^{(2)}-N\eta}+\omega_3^2
e^{-{2\eta\over3}+2\Theta-\chi^{(1)}-\chi^{(2)}+\chi^{(4)}-N\eta}f^{(-)}_1=0,\label{BAE-8}
\eea
where
\bea
\Theta=\sum_{l=1}^N\theta_l,\quad \chi^{(i)}=\sum_{l=1}^N\l^{(i)}_l,\quad i=1,2,3,4.\no
\eea

It is easy to check that the inhomogeneous $T-Q$ relations (\ref{T-Q-1}) and (\ref{T-Q-2}) fulfill the relations (\ref{Eigenvalue-function-1})-(\ref{Eigenvalue-function-3})
and the periodicity properties (\ref{Eigenvalue-function-4}).  The BEAs (\ref{BAE-5})-(\ref{BAE-8})
ensure that the inhomogeneous $T-Q$ relations (\ref{T-Q-1}) and (\ref{T-Q-2}) indeed
satisfy the asymptotic behavior (\ref{asym-su3-1})-(\ref{asym-su3-2}), while the BAEs (\ref{BAE-1})-(\ref{BAE-4}) assure that the inhomogeneous $T-Q$ relations have no
singularity at points $\l^{(i)}_l$.

\section{Results for the $su(n)$ spin torus}
\setcounter{equation}{0}

For the $su(n)$ case, by using the similar method introduced in the  previous section, we can derive that the fused transfer matrices $\{t_j(u)|j=1,\cdots, n\}$
satisfy the analogous operator product identities such as (\ref{trans-trans}), (\ref{trans-trans-1}), (\ref{fusion-orth}),
(\ref{Periodicity-T-1}) and (\ref{transfer-asym})-(\ref{transfer-asym-1}). These identities lead to that the corresponding eigenvalues
$\{\L_j(u)|j=1,\cdots,n\}$ satisfy the functional  relations:
\bea
&&\Lambda(\theta_j)\Lambda_m(\theta_j-\eta)=
\Lambda_{m+1}(\theta_j),\qquad  m=1,\cdots,n-1,\quad j=1,\cdots N,\label{Eigenvalue-n-1}\\[6pt]
&&\Lambda_m(\theta_j+k\eta)=0,\quad k=1,\cdots,m-1,\quad m=1,\cdots,n-1,\quad j=1,\cdots N,\label{Eigenvalue-n-2}\\[6pt]
&&\Lambda_n(u)=(-1)^{n-1}\prod^N_{l=1}\sinh(u-\theta_l+{\eta})\prod^{n-1}_{k=1}\sinh(u-\theta_l-k{\eta})\times {\rm id},\label{Eigenvalue-n-3}\\[6pt]
&&\Lambda_m(u+i\pi)=e^{-m({2\over n})i\pi}((-1)^N)^m\Lambda_m(u),\quad m=1,\cdots,n-1,\label{Eigenvalue-n-4}\\[6pt]
&&e^{-u+2({m\over n})u}\Lambda_m(u) \propto e^{\pm(mN-1)u}+\cdots,\quad u\rightarrow\pm\infty,\quad m=1,\cdots,n-1.\label{Eigenvalue-n-5}
\eea
Similar to the $su(3)$ case, the above relations completely determine the eigenvalues $\L_i(u)$
and thus enable us to express them in
terms of certain inhomogeneous $T-Q$ relations as those given by (\ref{T-Q-1})-(\ref{T-Q-2}). For the $su(n)$ case, let us
introduce the functions:
\bea
&&Q^{(i)}(u)=\prod_{l=1}^{N_i}\sinh(u-\l^{(i)}_l),\quad i=1,\cdots,2n-2,\label{Q-functions-n-1}\\[6pt]
&&Z_1(u)=e^{\phi_1}e^{(2-\frac{2}{n})u}a(u)\frac{Q^{(1)}(u-\eta)}{Q^{(2)}(u)},\no\\[6pt]
&&Z_2(u)=e^{\phi_2}\omega_n e^{-{2(u+\eta)\over{n}}} d(u)\frac{Q^{(2)}(u+\eta)Q^{(3)}(u-\eta)}{Q^{(1)}(u)Q^{(4)}(u)},\no\\[6pt]
&&\qquad\qquad\vdots\no\\[6pt]
&&Z_i(u)=e^{\phi_i}\omega_n^{i-1} e^{-{2(u+(i-1)\eta)\over{n}}} d(u)\frac{Q^{(2i-2)}(u+\eta)Q^{(2i-1)}(u-\eta)}{Q^{(2i-3)}(u)Q^{(2i)}(u)},
\no\\[6pt]
&&\qquad\qquad\vdots\no\\[6pt]
&&Z_{n-1}(u)=e^{-\sum_{j=1}^{n-2}\phi_j}\omega_n^{n-2} e^{-{2({u+(n-2)\eta})\over{n}}} d(u)\frac{Q^{(2n-4)}(u+\eta)Q^{(2n-3)}(u-\eta)}{Q^{(2n-5)}(u)Q^{(2n-2)}(u)},\no\\[6pt]
&&Z_n(u)=\omega_n^{n-1} e^{-{2({u+(n-1)\eta})\over{n}}} d(u)\frac{Q^{(2n-2)}(u+\eta)}{Q^{(2n-3)}(u)},\no
\eea
and
\bea
&&X_1(u)=e^{(1-\frac{2}{n})u}a(u)d(u)\frac{Q^{(3)}(u-\eta)f_1(u)}{Q^{(1)}(u)Q^{(2)}(u)},\no\\[6pt]
&&X_2(u)=e^{(1-\frac{2}{n})u}a(u)d(u)\frac{Q^{(2)}(u+\eta)Q^{(5)}(u-\eta)f_2(u)}{Q^{(3)}(u)Q^{(4)}(u)},\no\\[6pt]
&&\qquad\qquad\vdots\no\\[6pt]
&&X_i(u)=e^{(1-\frac{2}{n})u}a(u)d(u)\frac{Q^{(2i-2)}(u+\eta)Q^{(2i+1)}(u-\eta)f_i(u)}{Q^{(2i-1)}(u)Q^{(2i)}(u)},\no\\[6pt]
&&\qquad\qquad\vdots\no\\[6pt]
&&X_{n-1}(u)=e^{(1-\frac{2}{n})u}a(u)d(u)\frac{Q^{(2n-4)}(u+\eta)f_{n-1}(u)}{Q^{(2n-3)}(u)Q^{(2n-2)}(u)}.\label{X-function-n-1}
\eea
Here  $\omega_n=e^{\frac{2i\pi}{n}}$ such that $\omega_n^n=1$, the functions $\{f_i(u)|i=1,\cdots,n-1\}$ are
\bea
&&f_1(u)=f_1^{(+)}e^u+f_1^{(-)}e^{-u},\label{f-function-1}\\[6pt]
&&f_i(u)=f_i^{(-)}e^{-u},\quad i=2,\cdots,n-1.\label{f-function-2}
\eea
The $2(n-1)$ constants $f_1^{(\pm)}$, $\{f_i^{(-)}|i=2,\cdots,n-1\}$ and $\{\phi_i,i=1,\cdots,n-2\}$ are to be determined later. We define further functions $\{Y_l(u)|l=1,\cdots,2n-1\}$,
\bea
\lt\{\begin{array}{ll}Y_{2j-1}(u)=Z_j(u),& j=1,\cdots,n,\\[6pt]
Y_{2j}(u)=X_j(u),& j=1,\cdots,n-1,
\end{array}\rt.\label{Def-Y}
\eea
and take the notation
\bea
Y_j^{(l)}(u)=Y_j(u-l\eta),\quad l=1,\cdots,n,\quad j=1,\cdots,2n-1.\label{Notation-1}
\eea
The eigenvalue  $\{\Lambda_m(u)|m=1,\cdots n-1\}$ satisfying the relations (\ref{Eigenvalue-n-1})-(\ref{Eigenvalue-n-5})  can
be given in terms of the inhomogeneous $T-Q$ relations as
\footnote{For the $n=2$ case, the corresponding inhomogeneous $T-Q$ relation reduces to the alternative inhomogeneous $T-Q$ given in \cite{Wan15} (i.e., the equation (4.4.1) of subchapter 4.4).}
\bea
\Lambda_m(u)={\sum}'_{1\leq i_1<i_2<\cdots<i_m\leq 2n-1}Y_{i_1}(u)Y^{(1)}_{i_2}(u)\cdots Y^{(m-1)}_{i_m}(u),\quad
m=1,\cdots,n-1.\label{T-Q-n-1}
\eea
The sum $\sum'$ is over the constrained increasing sequences $1\leq i_1<i_2<\cdots<i_m\leq 2n-1$
such that when any $i_k=2j$ (i.e., $Y_{i_k}^{(k-1)}(u)=Y_{2j}^{(k-1)}(u)=X^{(k-1)}_j(u)$), then $i_{k-1}\leq 2j-3$  and $i_{k+1}\geq 2j+3$.
Namely, when $Y_{i_k}(u)=X_j(u)$,  the previous element $Y_{i_{k-1}}(u)$ and the next element $Y_{i_{k+1}}(u)$ can not be chosen as its nearest neighbors
(e.g., $X_{j-1}(u)$, $Z_j(u)$, $Z_{j+1}(u)$ and $X_{j+1}(u)$) in the diagram (\ref{struc-inhomo}).
\bea
\xymatrix{
Z_1\ar@{-}[rr]\ar@{-}[dr]&   &Z_2\ar@{-}[dr]\ar@{-}[rr]&   &Z_3\ar@{-}[dl]\ar@{-}[rr]&   &Z_4\ar@{-}[dl]\ar@{.}[rr]& &Z_{n}\ar@{.}[dl]\\
   &X_1\ar@{-}[rr] \ar@{-}[ur]&   &X_2\ar@{-}[rr]&   &X_3\ar@{-}[ul]\ar@{.}[rr]&   &X_{n-1}}
\label{struc-inhomo}\eea

To satisfy the asymptotic behaviors (\ref{Eigenvalue-n-5}), the non-negative integers
$\{N_i|i=1,\cdots,2(n-1)\}$ must be chosen as follows:
\begin{itemize}
\item For odd $n$,
\bea
N_{2i-1}=N_{2i}=N_{2(n-i)-1}=N_{2(n-i)}=\frac{i(n-i)}{2}N,\quad
i=1,\cdots,\frac{n-1}{2}.\label{N-1}
\eea

\item For even $n$ and even $N$,
\bea
N_{2i-1}=N_{2i}=N_{2(n-i)-1}=N_{2(n-i)}=\frac{i(n-i)}{2}N,\quad
i=1,\cdots,\frac{n}{2}.\label{N-2}
\eea

\item For even $n$ and odd $N$,
\bea
N_{2i-1}=N_{2i}=N_{2(n-i)-1}=N_{2(n-i)}=\frac{i(n-i)}{2}N+\frac{i}{2},\quad
i=1,\cdots,\frac{n}{2}.\label{N-3}
\eea
In contrast with (\ref{f-function-2}),  the function $f_{\frac{n}{2}}(u)$ now should be adjusted to
\bea
f_{n\over2}(u)=\sinh(u)\,f_{n\over2}^{(-)}\,e^{-u}.\label{f-function-3}
\eea
\end{itemize}
We present the details for the $su(4)$ case in Appendix B, which could be crucial to understand the structure for $n\geq4$.

It is easy to check that if $\{N_i|i=1,\cdots,2(n-1)\}$  are chosen as (\ref{N-1})-(\ref{N-3}) the corresponding inhomogeneous
$T-Q$ relations (\ref{T-Q-n-1}) have the asymptotic behavior
\bea
&&e^{-u+2({m\over n})u}\Lambda_m(u)=F^{(\pm)}_m\,e^{\pm(mN+1)u}+{F^{(\pm)}_m}'\,e^{\pm(mN-1)u}+\cdots, \quad u\rightarrow\pm\infty,\no\\[4pt]
&&\quad\quad m=1,\cdots,n-1.
\eea
The $2(n-1)$ coefficients $F^{(\pm)}_m$ are the functions of the parameters $\{\l^{(i)}_j|i=1,\cdots, 2(n-1);j=1,\cdots,N_i\}$,
$f_1^{(\pm)}$, $\{f_i^{(-)}|i=2,\cdots,n-1\}$ and $\{\phi_i,i=1,\cdots,n-2\}$. The explicit expressions can be obtained by direct
calculation. To make (\ref{Eigenvalue-n-5}) satisfied, the coefficients $F^{(\pm)}_m$ must vanish which leads to the $2(n-1)$
BAEs (for an example, (\ref{BAE-5})-(\ref{BAE-8}) for the $su(3)$ case)
\bea
F_m^{(\pm)}=0,\quad m=1,\cdots,n-1.\label{BAE-n-1}
\eea
Moreover, the vanishing condition of the residues of $\L_m(u)$ at the points $\l^{(i)}_j$ gives rise to the other BAEs:
\bea
&&e^{\phi_2}\omega_n e^{-\lambda^{(1)}_j-{2\eta\over{n}}} \frac{Q^{(2)}(\lambda^{(1)}_j+\eta)}{Q^{(4)}(\lambda^{(1)}_j)}
+a(\lambda^{(1)}_j)\frac{f_1(\lambda^{(1)}_j)}{Q^{(2)}(\lambda^{(1)}_j)}=0,\qquad j=1,\cdots,N_1,\\[6pt]
&&e^{\phi_1}e^{\lambda^{(2)}_j}Q^{(1)}(\lambda^{(2)}_j-\eta)
+d(\lambda^{(2)}_j)
\frac{Q^{(3)}(\lambda^{(2)}_j-\eta)f_1(\lambda^{(2)}_j)}{Q^{(1)}(\lambda^{(2)}_j)}=0,
\quad j=1,\cdots,N_2,
\eea
\bea
&&e^{\phi_i}\omega_n^{i-1}e^{-\lambda_j^{(2i)}-\frac{2(i-1)\eta}{n}}
\frac{Q^{(2i-1)}(\lambda_j^{(2i)}-\eta)}{Q^{(2i-3)}(\lambda_j^{(2i)})}
+a(\lambda_j^{(2i)})
\frac{Q^{(2i+1)}(\lambda_j^{(2i)}-\eta)f_i(\lambda_j^{(2i)})}{Q^{(2i-1)}(\lambda_j^{(2i)})}=0,\no\\[6pt]
&&\qquad\qquad i=2,\cdots,n-2,\quad j=1,\cdots,N_{2i},\\[6pt]
&&e^{\phi_{i+1}}\omega_n^{i}e^{-\lambda_j^{(2i-1)}-\frac{2i\eta}{n}}
\frac{Q^{(2i)}(\lambda_j^{(2i-1)}+\eta)}{Q^{(2i+2)}(\lambda_j^{(2i-1)})}
+a(\lambda_j^{(2i-1)})
\frac{Q^{(2i-2)}(\lambda_j^{(2i-1)}+\eta)f_i(\lambda_j^{(2i-1)})}{Q^{(2i)}(\lambda_j^{(2i-1)})}=0,\no\\[6pt]
&&\qquad \qquad i=2,\cdots,n-3,\quad j=1,\cdots,N_{2i-1},
\eea
\bea
&&e^{-\sum_{j=1}^{n-2}\phi_j}\omega_n^{n-2} e^{-\lambda_j^{(2n-5)}-{2(n-2)\eta\over{n}}} \frac{Q^{(2n-4)}(\lambda_j^{(2n-5)}+\eta)}{Q^{(2n-2)}(\lambda_j^{(2n-5)})}\no\\[6pt]
&&\qquad+a(\lambda_j^{(2n-5)})\frac{Q^{(2n-6)}(\lambda_j^{(2n-5)}+\eta)f_{n-2}(\lambda_j^{(2n-5)})}{Q^{(2n-4)}(\lambda_j^{(2n-5)})}=0,
\quad j=1,\cdots,N_{2n-5},
\eea
\bea
&&\omega_n^{n-1} e^{-\lambda_j^{(2n-3)}-{2(n-1)\eta\over{n}}} \hspace{-0.1truecm}Q^{(2n-2)}(\lambda_j^{(2n-3)}\hspace{-0.1truecm}+\hspace{-0.1truecm}\eta)
\hspace{-0.1truecm}+\hspace{-0.1truecm}a(\lambda_j^{(2n-3)})
\frac{Q^{(2n-4)}(\lambda_j^{(2n-3)}+\eta)f_{n-1}(\lambda_j^{(2n-3)})}{Q^{(2n-2)}(\lambda_j^{(2n-3)})}=0,\no\\[6pt]
&&\qquad\quad\quad\qquad j=1,\cdots,N_{2n-3},\quad\\[6pt]
&&e^{-\sum_{j=1}^{n-2}\phi_j}\omega_n^{n-2} e^{-\lambda_j^{(2n-2)}-{2(n-2)\eta\over{n}}} \frac{Q^{(2n-3)}(\lambda_j^{(2n-2)}-\eta)}{Q^{(2n-5)}(\lambda_j^{(2n-2)})}
+a(\lambda_j^{(2n-2)})\frac{f_{n-1}(\lambda_j^{(2n-2)})}{Q^{(2n-3)}(\lambda_j^{(2n-2)})}=0,
\no\\[6pt]
&&\qquad\quad\quad\qquad j=1,\cdots,N_{2n-2}.\label{BAE-n-2}
\eea
Associated with the BAEs (\ref{BAE-n-1})-(\ref{BAE-n-2}), the inhomogeneous $T-Q$ relation (\ref{T-Q-n-1}) give the eigenvalues of the transfer matrix of the $su(n)$ spin torus.


\section{Conclusions}

In this paper, we have studied the $su(n)$ spin torus described by the Hamiltonian (\ref{Ham}) with
the anti-periodic boundary condition (\ref{anti-boundary}). In the framework of  ODBA, we have obtained the
eigenvalues of the corresponding transfer matrix in terms of the inhomogeneous $T-Q$ relation (\ref{T-Q-n-1})
and  the associated BAEs (\ref{BAE-n-1})-(\ref{BAE-n-2}). The exact spectrum obtained in this paper allows us further to construct the corresponding eigenstates.
The results will be presented elsewhere \cite{Hao16}.

\section*{Acknowledgments}

The financial supports from the National Natural Science Foundation
of China (Grant Nos. 11375141, 11374334, 11434013, 11425522 and 11547045), BCMIIS and the Strategic Priority Research Program
of the Chinese Academy of Sciences are gratefully acknowledged.

\section*{Appendix A: Generic twist matrices}
\setcounter{equation}{0}
\renewcommand{\theequation}{A.\arabic{equation}}
A generic twist matrix ${\mathcal{G}}$ associated with an integrable boundary satisfies the relation
\bea
R_{12}(u-v)\,{\mathcal{G}}_1\,{\mathcal{G}}_2={\mathcal{G}}_1\,{\mathcal{G}}_2\,R_{12}(u-v).\label{Invariant-R-1}
\eea
Normally, ${\mathcal{G}}$ is a c-number matrix. The solutions of (\ref{Invariant-R-1}) with the $R$-matrix given by (\ref{R-matrix-1}) can be specified as
$n$ classes labeled by $l$
\bea
{\mathcal{D}}\,g^l,\quad l=0,\cdots,n-1,\label{Twist-matrices}
\eea
where ${\mathcal{D}}$ is an arbitrary non-degenerate diagonal $n\times n$ matrix and $g$ is given by (\ref{g-matrix}). All solutions to (\ref{Invariant-R-1}) are
some  products of elements of these classes.   ${\mathcal{G}}={\mathcal{D}}$
corresponds to the diagonal twisted boundary condition (including the periodic boundary condition ${\mathcal{D}}={\rm id}$ as a special case).
Without losing the generality, in this paper we consider the twist matrix ${\mathcal{G}}=g$ which corresponds to the antiperiodic boundary condition (\ref{anti-boundary}). The generalization to the other cases is
straightforward.

\section*{Appendix B: $T-Q$ relation for the $su(4)$ spin torus}
\setcounter{equation}{0}
\renewcommand{\theequation}{B.\arabic{equation}}
For $n=4$, the functions (\ref{Q-functions-n-1})-(\ref{X-function-n-1}) and the functions $X_j(u)$ read
\bea
&&Q^{(i)}(u)=\prod_{l=1}^{N_i}\sinh(u-\l^{(i)}_l),\quad i=1,\cdots,6,\\[6pt]
&&Z_1(u)=e^{\phi_1}e^{\frac{3u}{2}}a(u)\frac{Q^{(1)}(u-\eta)}{Q^{(2)}(u)},\no\\[6pt]
&&Z_2(u)=e^{\phi_2}\omega_4 e^{-{u+\eta\over2}} d(u)\frac{Q^{(2)}(u+\eta)Q^{(3)}(u-\eta)}{Q^{(1)}(u)Q^{(4)}(u)},\no\\[6pt]
&&Z_3(u)=e^{-\phi_1-\phi_2}\omega_4^2 e^{-{u\over2}-\eta} d(u)\frac{Q^{(4)}(u+\eta)Q^{(5)}(u-\eta)}{Q^{(3)}(u)Q^{(6)}(u)},\no\\[6pt]
&&Z_4(u)=\omega_4^3 e^{-{{u+3\eta}\over2}} d(u)\frac{Q^{(6)}(u+\eta)}{Q^{(5)}(u)},\no\\[6pt]
&&X_1(u)=e^{\frac{u}{2}}a(u)d(u)\frac{Q^{(3)}(u-\eta)f_1(u)}{Q^{(1)}(u)Q^{(2)}(u)},\no\\[6pt]
&&X_2(u)=e^{\frac{u}{2}}a(u)d(u)\frac{Q^{(2)}(u+\eta)Q^{(5)}(u-\eta)f_2(u)}{Q^{(3)}(u)Q^{(4)}(u)},\no\\[6pt]
&&X_3(u)=e^{\frac{u}{2}}a(u)d(u)\frac{Q^{(4)}(u+\eta)f_3(u)}{Q^{(5)}(u)Q^{(6)}(u)}.
\eea
Here $\omega_4=e^{\frac{2i\pi}{4}}$. The corresponding $T-Q$ relations (\ref{T-Q-n-1}) become
\bea
&&\hspace{-1.2truecm}\L(u)=Z_1(u)+Z_2(u)+Z_3(u)+Z_4(u)+X_1(u)+X_2(u)+X_3(u),\\[6pt]
&&\hspace{-1.2truecm}\L_2(u)=Z_1(u)Z_2(u-\eta)+Z_1(u)X_2(u-\eta)+Z_1(u)Z_3(u-\eta)+Z_1(u)X_3(u-\eta)\no\\[6pt]
&&+Z_1(u)Z_4(u-\eta)+X_1(u)Z_3(u-\eta)+X_1(u)X_3(u-\eta)+X_1(u)Z_4(u-\eta)
\no\\[6pt]
&&+Z_2(u)Z_3(u-\eta)+Z_2(u)X_3(u-\eta)+Z_2(u)Z_4(u-\eta)+X_2(u)Z_4(u-\eta)\no\\[6pt]
&&+Z_3(u)Z_4(u-\eta),\\[6pt]
&&\hspace{-1.2truecm}\L_3(u)=Z_1(u)Z_2(u-\eta)Z_3(u-2\eta)+Z_1(u)Z_2(u-\eta)X_3(u-2\eta)\no\\[6pt]
&&+Z_1(u)Z_2(u-\eta)Z_4(u-2\eta)+Z_1(u)X_2(u-\eta)Z_4(u-2\eta)\no\\[6pt]
&&+Z_1(u)Z_3(u-\eta)Z_4(u-2\eta)+X_1(u)Z_3(u-\eta)Z_4(u-2\eta)\no\\[6pt]
&&+Z_2(u)Z_3(u-\eta)Z_4(u-2\eta),\\[6pt]
&&\hspace{-1.2truecm}\Lambda_4(u)=Z_1(u)Z_2(u-\eta)Z_3(u-2\eta)Z_4(u-3\eta).
\eea
\begin{itemize}
\item For even $N$ ,
\bea
N_1=N_2=N_5=N_6=\frac32 N, \quad N_3=N_4=2N,
\eea
and the functions $f_i(u)$ are :
\bea
f_1(u)=f_1^{(+)}e^u+f_1^{(-)}e^{-u},\, f_2(u)=f_2^{(-)}e^{-u},\,
f_3(u)=f_3^{(-)}e^{-u}.
\eea

\item For odd $N$,
\bea N_1=N_2=N_5=N_6=\frac{3N+1}2, \quad
N_3=N_4=2N+1,
\eea
and the functions $f_i(u)$ are :
\bea
f_1(u)=f_1^{(+)}e^u+f_1^{(-)}e^{-u},\, f_2(u)=\sinh(u)\,f_2^{(-)}\,e^{-u},\,
f_3(u)=f_3^{(-)}e^{-u}.
\eea
\end{itemize}

\end{document}